# Study on Neural Immune PD Type Tracking Control for DC Actuating Mechanism

YongChol Sin, HyeGyong Sin, GyongIl Ryang,

**Kim Il Sung** University, Pyongyang

Democratic People's Republic of Korea

**Abstract:** Artificial Immune Systems(AIS) have been widely used in different fields, such as control, robotics, computer science and multi-agent systems.

In this paper is proposed a new approach of neural immune PD type tracking control combining artificial immune control with neural network.

It is assumed that the output of the helper T-cell is concerned with not only the error of system but also its changing rate, while the output of suppressor T-cell is unknown nonlinear function with respect to the amount and changing rate of antigens and the changing rate of antibodies, which is approximated by the output of neural network. From this, we derive neural immune PD type control law and apply it to the trajectory tracking of DC actuating mechanism.

The validity of the proposed method is verified by simulation and the simulation results show that this method can follow the desired trajectory more rapidly and more accurately compared to the previous method.

**Keywords:** Artificial Immune Systems(AIS), immune control, neural immune PD type control,

## Ⅰ. Introduction

The immune system is one of the four nature information processing systems, including brain neural system, genetic system and incretion system, in biological system.

Immune system has the ability to defend system against foreign material and the relation between immune factors is differently modeled. Many attempts have been done to apply the above ability of the biological immune to the fields of control and a new control approach, called artificial immune control, has been developed.[1~3]

The focus of study on artificial immune control strategies has been concentrated on the solution of the learning problems and adaptability problems in the control system by using immune mechanisms. Many applications using artificial immune control such as immune P controller and immune PID controller have been developed.[1] The intelligent control methods combining immune control technique with soft-computing technique have been also proposed.[5-7]

In most literature, the output of helper T-cell is considered as the term related only to the error of system, while the output of suppressor T-cell is determined as different types of functions. (where the system error and the control input are treated as antigen and antibody respectively.) [4] But in biological immune system the production of antibodies could be affected by the increasing or decreasing rate as well as the amount of antigens. From this point of view, in this paper a new control method, neural immune PD type control approach, is proposed and applied to the trajectory tracking of DC actuating mechanism.

In Section Ⅱ, the neural immune PD type control system is designed and in Section Ⅲ the neural immune PD type tracking control law for DC actuating mechanism is derived. Section Ⅳ considers the configuration of the neural controller and, finally, Section Ⅴ presents simulation results for the trajectory tracking control.

## Ⅱ. Design of the Neural Immune PD Type Control System

The general immune control law is presented as following

$$u(t) = P_h(e(t))[1 - f_h(e(t)) \cdot f_s(\dot{u}(t))] \quad (1)$$

where $e(t)$ is a system error, $P_h(\cdot)$, $f_s(\cdot)$ are functions expressing the outputs of the helper T-cell and suppressor T-cell respectively and $f_h(\cdot)$ is the inverse function of $P_h(\cdot)$.[1]

Based on the biological immune ability, we assume to get more similar model of immune systems as following:

*[Assumption 1]* The output of helper T-cell is related not only to the amount of antigens but to the changing rate.

*[Assumption 2]* The output of the suppressor T-cell is the unknown nonlinear function with respect to $e(t), \dot{e}(t), \dot{u}(t)$.

From these assumptions, we determine the output of helper T-cell as

$$u_h(t) = P_h(e(t), \dot{e}(t)) = K_P e(t) + K_D \dot{e}(t) \quad (2)$$

where $e(t)$ is the error between the desired trajectory and the actual output of the plant.

The output of the suppressor T-cell is as following

$$u_s(t) = f(e(t), \dot{e}(t), \dot{u}(t)) \quad (3)$$

and the resulting law of artificial immune PD type control is rewritten as

$$\begin{aligned}u(t) &= P_h(t)[1 - f_h(t) \cdot f_s(t)] \\ &= K_P e(t) + K_D \dot{e}(t) - f(e(t), \dot{e}(t), \dot{u}(t))\end{aligned} \quad (4)$$

We try to approximate $u_s(t)$ by using the output of the neural network as

$$u_s(t) = f_N(e(t), \dot{e}(t), \dot{u}(t)) \quad (5)$$

where function $f_N$ expresses the output of neural network.

Consequently Eq. (4) is rewritten as

$$\begin{aligned}u(t) &= u_h(t) - u_s(t) \\ &= K_P e(t) + K_D \dot{e}(t) - f_N(e(t), \dot{e}(t), \dot{u}(t))\end{aligned} \quad (6)$$

*[Definition 1]* The artificial immune control, in which the output of helper T-cell is constructed in the form of PD and the output of suppressor T-cell is obtained by neural network as Eq. (5), will be called *neural immune PD type control*.

The corresponding block diagram of neural immune PD type control system is shown in Fig.1.

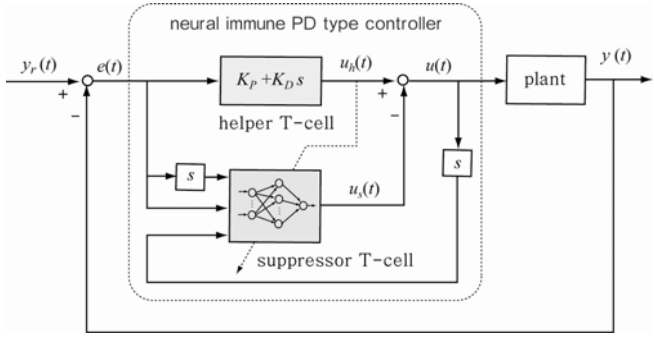

Fig.1. Block diagram of neural immune PD type control system

*[Definition 2]* The neural immune PD type control to track the desired second-order differentialable desired trajectory $\theta_d(t)$ will be called *neural immune PD type tracking control*.

In the next section, neural immune PD type tracking control of DC actuating mechanism combining DC motor with load link will be considered.

## Ⅲ. Neural Immune PD Type Tracking Control of DC actuating mechanism

The DC actuating mechanism, in which load is linked with DC motor, consists of the electric circuit part and the mechanical part as shown in Fig. 2.

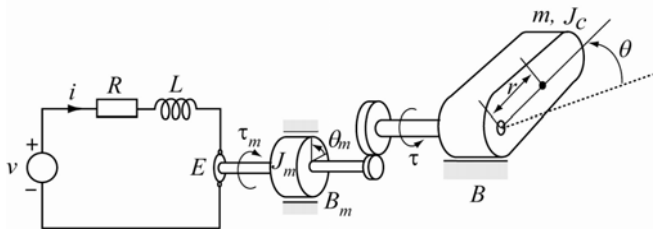

Fig.2. DC Actuating Mechanism

The mechanical part consists of load link and armature link and the motion equation of the load link is presented as

$$(J_c + mr^2)\ddot{\theta} + B\dot{\theta} + mgr\cos\theta = \tau \qquad (7)$$

and the equation of the armature link is expressed as

$$\tau_m = J_m\ddot{\theta}_m + B_m\dot{\theta}_m + \tau_{sm} = J_m\ddot{\theta}_m + B_m\dot{\theta}_m + \frac{\tau}{j} \qquad (8)$$

From above two equations, the resulting motion equation of the mechanical part is rewritten as

$$\begin{aligned}
\tau_m &= \left(J_m + \frac{J}{j^2}\right)\ddot{\theta}_m + \left(B_m + \frac{B}{j^2}\right)\dot{\theta}_m + \frac{1}{j}mgr\cos(\theta_m/j) \\
&= J_e\ddot{\theta}_m + B_e\dot{\theta}_m + \frac{1}{j}mgr\cos(\theta_m/j) \\
&= j(J_e\ddot{\theta} + B_e\dot{\theta}) + \frac{1}{j}mgr\cos(\theta)
\end{aligned} \qquad (9)$$

On other hand, the voltage-current relationship of the armature circuit is expressed as

$$v = iR + L\frac{di}{dt} + E \qquad (10)$$

The dynamic model of the total system combining the electric circuit part with mechanical part is simply expressed as

$$a_2\ddot{\theta}(t) + a_1\dot{\theta}(t) + a_0\cos\theta(t) = v(t) \qquad (11)$$

, where parameters $a_2, a_1, a_0$ are

$$a_2 = j\frac{J_e R}{k_t}, \ a_1 = j\frac{B_e R + k_v k_t}{k_t}, \ a_0 = \frac{R}{jk_t}mgr$$

Then the linear nominal model for Eq. (11) can be described as

$$\hat{a}_2\ddot{\theta}(t) + \hat{a}_1\dot{\theta}(t) = v(t) \qquad (12)$$

, where parameters $\hat{a}_2, \hat{a}_1$ are the nominal values of parameters $a_2, a_1$

Considering its nominal model (12), the dynamic model (11) can be rewritten as

$$\hat{a}_2\ddot{\theta}(t) + \hat{a}_1\dot{\theta}(t) + d(\ddot{\theta}, \dot{\theta}, \theta) = v(t) \qquad (13)$$

, where

$$d(\ddot{\theta}, \dot{\theta}, \theta) = \Delta a_2\ddot{\theta}(t) + \Delta a_1\dot{\theta}(t) + a_0\cos\theta(t) \qquad (14)$$

is *equivalent disturbance* acted upon the system.

Based on the above neural immune PD type control law, we design the tracking control law as

$$v(t) = v_h(t) - v_s(t)$$
$$= \hat{a}_2[\ddot{\theta}_d(t) + K_D\dot{e}(t) + K_P e(t)] + \hat{a}_1\dot{\theta}(t) - f_N(e(t), \dot{e}(t), \dot{v}(t))$$
(15)

, where

$$v_h(t) = \hat{a}_2[\ddot{\theta}_d(t) + K_D\dot{e}(t) + K_P e(t)] + \hat{a}_1\dot{\theta}(t)$$
(16)

is the output of helper T-cell and

$$v_s(t) = f_N(e(t), \dot{e}(t), \dot{v}(t))$$ (17)

is the output of suppressor T-cell.

The feedback coefficients $K_P$, $K_D$ influence the dynamics of the closed-loop system as well as the convergence of position error and speed error and the neural network suppresses the equivalent disturbance $d(\ddot{\theta}, \dot{\theta}, \theta)$.

The block diagram of neural immune PD type tracking control system proposed above is shown in Fig.3.

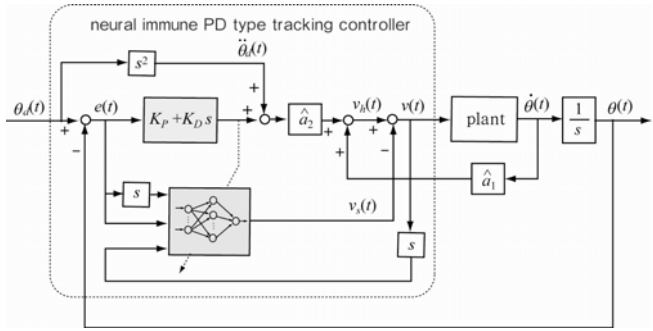

Fig.3. Block diagram of neural immune PD type tracking control system

In the DC actuating mechanism whose dynamics is given as the above Eq. (13), if the output of the neural controller $v_s(t) = f_N(e(t), \dot{e}(t), \dot{v}(t))$ accurately estimates and compensates $d(\ddot{\theta}, \dot{\theta}, \theta)$ (that is, suppresses), the dynamics of the closed-loop system is expressed as

$$\hat{a}_2[\ddot{e}(t) + K_D\dot{e}(t) + K_P e(t)] = 0$$ (18)

and if the feedback coefficients corresponding to the control input of the helper T-cells Eq.(16) satisfy the relation

$$K_D^2 = 4K_P \quad (K_P > 0, K_D > 0),$$ (19)

the desired dynamics of the closed-loop system and the exponential convergence of the tracking error will be achieved.

## IV. Design of the Neural Controller and Learning Algorithm

The recurrent neural network is used to generate the output of suppressor T-cell, where inputs and outputs of neural cells in the input, hiden and output layers (input cells 3, output cells 1) are determinated as

$$s_j(k+1) = \begin{cases} u_j(k+1), & (j \in I) \\ \sum_{i=1}^{N} w_{ji} x_i(k), & (j \in Q) \end{cases}$$ (20)

$$x_j(k+1) = \begin{cases} s_j(k+1), & (j \in I) \\ f_j[s_j(k+1)], & (j \in Q) \end{cases}$$ (21)

$$Q = H \cup O$$

, where $N = 4 + p$ is the total number of cells.

The activation function is defined as

$$f_j[s_j(k+1)] = \begin{cases} \dfrac{1-\exp[-Ts_j(k+1)]}{1+\exp[-Ts_j(k+1)]}, & (j \in H) \\ as_j(k+1), & (j \in O) \end{cases}$$ (22)

and then we try to define the cost function is as

$$J(k_0, k_f) = \frac{1}{2} \sum_{k=k_0+1}^{k_f} E^2(k)$$ (23)

, where
$$E(k) = K_D\dot{e}(k) + K_P e(k).$$

**[Learning Algorithm]**

**Step 1**: Design the structure of the recurrent neural network and initialize parameter values (initial weights, biases, learning rates and etc).

**Step 2**: At the discrete time $k = k_0, k_0+1, \cdots, k_f - 1$, $\dot{v}(k), e(k), \dot{e}(k)$ are fed to the neural network as input and the outputs of the cells are calculated according to Eq. (20)~Eq. (22).

**Step 3:** At the discrete time $k = k_0+1, k_0+2, \cdots, k_f$ the error between desired trajectory and actual trajectory $e(k) = \theta_d(k) - \theta(k)$ is calculated.

**Step 4**: Calculate $\delta(t_f)$ and $\delta(k)$ through the back direction of time (that is, $k_f \to k_0 + 1$) according to Eq. (24).

$$\delta(k) = \begin{cases} \begin{bmatrix} \mathbf{0}_p \\ \cdots\cdots\cdots\cdots \\ (f^O)'[K_D\dot{e}(k_f)+K_P e(k_f)] \end{bmatrix} = \begin{bmatrix} \mathbf{0}_p \\ \cdots\cdots \\ \delta^O(k_f) \end{bmatrix}, & (k=k_f) \\ \mathbf{F}\begin{bmatrix} \mathbf{0}_p \\ \cdots\cdots\cdots\cdots \\ K_D\dot{e}(k)+K_P e(k) \end{bmatrix} + \mathbf{W}^T\delta(k+1), & (k_0+1 \le k < k_f) \end{cases}$$

(24)

**Step 5**: Update weights according to learning rules Eq. (25)

$$w(l+1) = w(l) - \eta \sum_{k=k_0}^{k_f-1}\left[\frac{\partial s(k+1)}{\partial w(k)}\right]^T \delta(k+1). \quad (25)$$

**Step 6**: Iterate Step 2~Step 5 until learning converges.

## Ⅴ. Simulation

The validity of the proposed method is verified by simulation under the parameters $a_2 = 7.6$, $a_1 = 0.0234$, $a_0 = 0.26$, $K_P$=100, $K_D$=20.

The characteristics of tracking progress for desired trajectory and error convergence is shown in Fig.4.

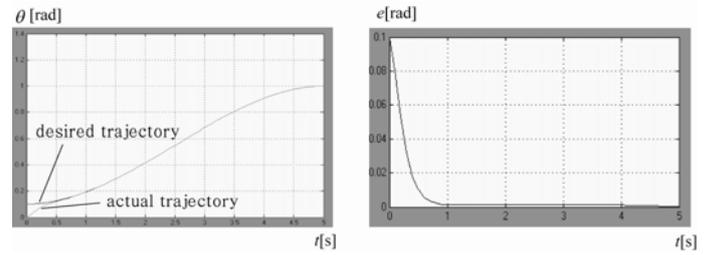

a) Tracking of desired trajectory  b) Error convergence

Fig.4. Simulation results of neural immune PD type tracking control

Tracking performance and error convergence results are shown in Fig. 5 according to different values of feedback coefficients $K_P$, $K_D$.

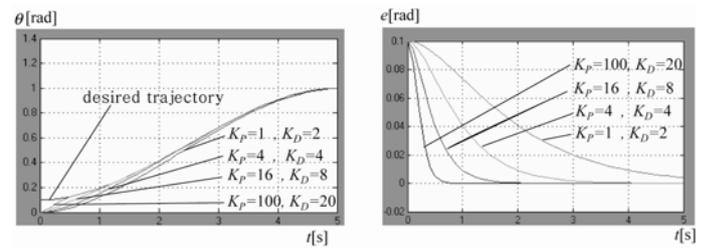

a) Tracking of desired trajectory  b) Error convergence

Fig.5. Tracking progress and error convergence according to $K_P$, $K_D$

Fig.6 shows the comparison of the proposed method with the previous one immune PID control method.

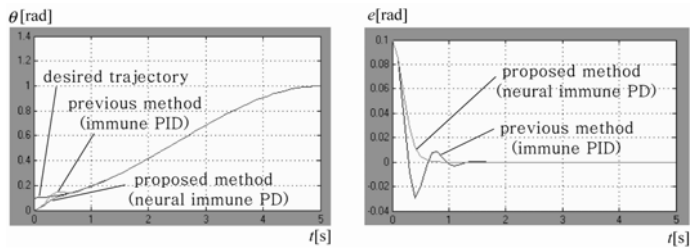

a) Tracking of desired trajectory    b) Error convergence

Fig.6. Compared results

## Ⅵ. Conclusions

In this paper, a new neural immune PD type control approach combining immune control law with neural network is proposed and applied to the trajectory tracking of DC actuating mechanism.

Simulation results show that the proposed method can achieve much better tracking of trajectory compared to the the previous immune PID control approach.

## References

[1]. Fu. Dongmei, Research of Immune Controller, 《Handbook of Research on Artificial Immune Systems and Natural Computing: Applying Complex Adaptive Technologies》, Medical Information Science Reference, New York, 262-304, 2009.

[2]. Lingli Yu, Zixing Cai, Zhoggyang Jiang, Qiang Hu, 《An Advanced Fuzzy Immune PID-type Tracking Controller of a Nonholonomic Mobile Robot》, Proceedings of the IEEE International Conference on Automation and Logistics, August, 18-21, 2007, Jinan, China.

[3]. Guan-Chun Luh, Wei-Ewn Liu, 《Reactive Immune Network Based Mobile Robot Navigation》, Artificial Immune Systems, Third International Conference, ICARIS 2004, 119-132, 2004.

[4]. Motohiro Kawafuku, Minoru Sasaki, KazuhikonTakahash, 《Adaptive Learning Method of Neural Network Controller using an Immune Feedback Law》, Proceedings of the 1999 IEEE/ASME International Conference on Advanced Intelligent Mechatronics, September 19-23, 1999, Atlanta, USA.

[5]. Takahashi K., Yamada T., 《A Self-tuning Immune Feedback Controller for Controlling Mechanical Systems》, AdvancedIntelligent Mechatronics '97, IEEE/ASME International Conference on, pp. 101, June 1997.

[6]. Yang Jian-guo, LI bei-zhi, Xiang Qian, 《Immune Genetic Algorithm for Optimal Design》, Jurnal of Dong Hua University, Vol.19, No.4, pp.16-19, 2002

[7]. Albert Ko, H. Y. K. Lau, T. L. Lau, 《An Immune Control Framework for Decentrolized Mechatronic Control》, Artificial Immune Systems, Third International Conference, ICARIS 2004, 91-105,2004.